\documentclass[a4]{article}

\def\Be{\begin{equation}}
\def\Ee{\end{equation}}
\def\Bea{\begin{eqnarray}}
\def\Eea{\end{eqnarray}}

\def\ee{\mbox{e}}

\def\ultm{{\rm{ult}}}
\def\ult{\beta}
\def\ordu{{\rm{ord}_{\beta}}}
\def\ordp{{\rm{ord}_{p}}}
\def\QQ{{\bf Q}}
\def\RR{{\bf R}}
\def\ZZ{{\bf Z}}

\begin{document}

{
\begin{center}
{\bf \Large Lotka-Volterra Equation over a Finite Ring $\ZZ/p^N\ZZ$}\\
{~} \\
{Shigeki Matsutani} \\
{8-21-1 Higashi-Linkan Sagamihara 228-0811 JAPAN} \\
{~}
\end{center}

\centerline{\Large \bf Abstract}

 Discrete Lotka-Volterra equation over $p$-adic space was constructed
 since $p$-adic space is a prototype of spaces with the non-Archimedean
 valuations and the space given by taking ultra-discrete limit studied
 in soliton theory should be regarded as a space with the
 non-Archimedean valuations in the previous report (solv-int/9906011).
In this article, using the natural projection from $p$-adic
 integer to a ring $\ZZ/p^n \ZZ$, a soliton equation is defined over
 the ring. Numerical computations shows that it behaves regularly.
\vskip 1.0 cm

\centerline{\Large \bf \S 1. Introduction}

\vskip 0.5 cm

According to [LL], studies on the ultrametric space which is
characterized by the strong triangle axiom
\Be
        d(x,z) \le \max[d(x,y),d(y,z)] 
\Ee
or for one-dimensional additive space case
\Be
        |x-z|_{\ultm} \leq \max[|x|_{\ultm},|z|_{\ultm}] 
\Ee
has been current for this 10-15 years in the fields of general
topology, computer language, rings of meromorphic function and so on.
This space is called  non-Archimedean space in English and German
literature and is known as ultrametric space in France literature and
as isosceles space in Russian [LL]. This space appeared, in first, in
the theory of number theory as $p$-adic (Hensel) integer but nowadays,
it is known that this space is natural even for the fields out of
number theory.

In fact, the space obtained by ultra-discrete limit which has been
studied in  soliton theory [TS, TTMS] holds the relation (1)  as shown
in [M]. (If one recognizes that the soliton theory is roughly a theory
of functions over a compact Riemann surface, the functions are also
governed by the non-Archimedean valuation [I] and thus the ultrametric
is built in soliton theory. Sato theory is based on the fact [S].)

In the studies of the ultrametric space, it is natural to  feel that
the theory over a field with characteristics $q=0$ is too restricted
because $p$-adic space is a prototype of the ultrametric space.

Actually as the ultra-discrete equation [TS, TTMS] is given as a
difference-difference equation, difference-difference equations are in
general given by  algebraic relations. Thus it is natural to consider
the equations in the algebraic category and there non-vanishing
characteristics is canonical. For example, the one-dimensional linear
difference-difference heat equation is algebraically defined as
follows: Let $K$ be a field with characteristics $q=0$ and $K[X,T]$ be
a set of polynomial of $X$ and $T$. Let us introduce a ring
$F_X[T]:=K[X,T]/(1-X^n)$ where $(1-X^n)$ is an ideal generated by
$1-X^n$ and $n$ is a positive integer. We have its subset
\Be
        F_X^i[T]:=\{
        f\in F_X[T]\ | \ \mbox{order of } f \mbox{ in } T
        \le i\ \}. 
\Ee
Then we can define a map $\phi:F^i_X[T]/F^{i-1}_X[T]\to
F^{i+1}_X[T]/F^{i}_X[T]$ by for an element $f_i$ of \break
$F^i[X,T]/F^{i-1}[X,T]$, $\phi f_i:=T(X-2+\epsilon +X^{-1})f_i$. For
an element $f_0\in F^0_X[X,T]$, we compute $f_n:=\phi^n f_0$. This is
the computation of the one-dimensional discrete heat equation with a
periodic boundary condition on $n$; $X$ and $T$ are shift operators.
Since $F_X^0[T]$ can be regarded as a cyclotomic field, one may compare
the norm of $|X-2+\epsilon +X^{-1}|$ with the parameter $\epsilon$,
which is related to stability's criterion of the numerical computation
[PTVF]. In this formulation, it is not strange to consider it over a
field with non-vanishing characteristics.

Similarly we wish to formulate the difference-difference non-linear
equation over a more general field  with non-vanishing characteristics
or its related ring. In fact, there were several attempts to formulate
the soliton theory over finite fields [N, NM]. The purpose of this
study including the previous report [M] is to extend the soliton theory
over $\RR$ to $p$-adic number space in order to consider the meanings
of the ultra-discrete limit. (As the $p$-adic valuation is a prototype
of the non-Archimedean valuation and the ultra-discrete system is
natural in soliton theory, it is very natural to investigate soliton
equations in the $p$-adic space.) The extension of soliton theory to
$p$-adic space has been done by Ichikawa for continuous soliton theory
or KP-hierarchy [Ic1-3]. In this study, we  restrict ourselves only to
consider the difference-difference equations. Then as mentioned in [M],
we can also define the $p$-adic difference-difference Lotka-Volterra
equation and show that its $p$-adic valuation version has the same
structure of the ultra-discrete difference-difference Lotka-Volterra
equation. This implies that there might exist a functor between
categories of discrete ordinary soliton equations and $p$-adic soliton
equations, even though we did not mention it [M].

In this article, we will investigate $p$-adic system more concretely.
Since $p$-adic integer is canonically connected with a finite ring
$\ZZ /p^N \ZZ$, we will consider that the Lotka-Volterra equation over
the finite ring $\ZZ /p^N \ZZ$. Due to its finiteness, we can compute
it concretely. Numerical computations show that  the Lotka-Volterra
equation over the finite ring $\ZZ /p^n \ZZ$ behaves regularly. It is
expected that there is a natural group governing the system.

In this article, we denote the set of integers, rational number
and real numbers by $\ZZ$, $\QQ$ and $\RR$ respectively.

\vskip 0.5 cm

{\centerline{\Large{\bf \S 2. Ultra-Discrete Limit as a Valuation }}}

\vskip 0.5 cm

This section reviews the previous report [M] briefly in order to
connect the ultra-discrete system with ultrametric system. Let
$\overline{{\cal A}_{[\beta]}}$ be a set of non-negative real valued
functions over $\{\beta \in \RR_{>0}\}$ where $\RR_{>0}$ is a set of
positive real numbers. Let us define a map $\ordu:\overline{{\cal
A}_{[\beta]}}\cup\{ {0} \} \to \RR+\infty$. We set $ \ordu(0) = \infty$
for $u\equiv0$ and for $u \in \overline{{\cal A}_{[\beta]}}$,
\Be
        \ordu(u):=
        -
        \lim_{ \beta \to +\infty} \frac{1}{\beta}\log ( u). 
\Ee
We call this value ultra-discrete of $u$.

Let us choose a subset ${\cal A}_{[\beta]}$ of $\overline{{\cal
A}_{[\beta]}}$, ${\cal A}_{[\beta]}:=\{ {u} \in \overline{{\cal
A}_{[\beta]}} \ | \ \ordu({u})< \infty\ \}$. Further we  identify the
set $\{ {u} \in \overline{{\cal A}_{[\beta]}}\ | \ \ordu({u})=\infty\
\}$ with $\{0\}$. The ultra-discrete $\ordu$ is a non-Archimedean
valuation [I] since it hold the properties (I${}_\beta$):

\vskip 0.5 cm

{\large \bf Proposition I${}_\beta$ }
{\it
For
$u, v \in {\cal A}_{[\beta]}\cup\{ {0} \}$,
\begin{enumerate}

\item $ \ordu(u v ) = \ordu(u) + \ordu( v ) $.

\item $ \ordu(u + v ) =\min( \ordu(u) , \ordu( v ))$.

\vskip 0.5 cm

\end{enumerate}
}

Let us, now, give the difference-difference Lotka-Volterra equation for
$\{ c^m_n\in \RR_{\ge 0}\ | \ (n,m) \in \Omega\times \ZZ\ \}$ [HT],
\Be
 \frac{c^{m+1}_n}{ c^m_n} = \frac{ 1 + \delta
 c^m_{n-1}} { 1 + \delta c^{m+1}_{n+1} }.
\Ee
Here $\Omega$ is a subset of $\ZZ$, $\delta$ is a small parameter
($|\delta|<1$) connecting between discrete system and continuum system,
$n$ is an index of a subset of the integer $\ZZ$ and $m$ is
of time step.

By introducing new variables $f^m_n := - \ordu( c^m_n )$ and
$d := -\ordu( \delta )$ [T], we have a ultra-discrete version of the
difference-difference Lotka-Volterra equation (5) for
$c^m_n\in {\cal A}_{[\beta]}$ and $\delta \in {\cal A}_{[\beta]}$
[TS, T, TTMS],
\Be
f^{m+1}_n - f^m_n = \ordu( 1 + \delta_p c^m_{n-1})
- \ordu(1 + \delta_p c^m_{n-1}).
\Ee
We emphasize that (6) is considered as a valuation version
of the difference-difference soliton equation (5).

Now in order to connect the ultra-discrete valuation and ultrametric
in the framework [LL], we introduce a real number   $\overline \beta
\gg 1$ and define a quantity  for $x \in {\cal A}_{[\beta]}$ as,
\Be
        |x|_{\ult} := \left
        (\ee^{ -\bar\beta}\right)^{\ordu(x)}. 
\Ee
This is an ultrametric because it satisfies next proposition.

\vskip 0.5 cm

{\large \bf Proposition II${}_\beta$} {\it
For $u$, $v \in {\cal A}_{[\beta]}\cup \{ {0} \}$,
$|u|_{\ult}$ and $|v|_{\ult}$ hold following properties,

\begin{enumerate}

\item $|u|_{\ult}$ depends upon $\bar \beta$.

\item if $ |v|_{\ult} =0 $, $v=0.$

\item $ |v|_{\ult} \ge0 $.

\item $ |v u |_{\ult}= |v|_{\ult} |u|_{\ult} $.

\item $|u + v|_{\ult} = \max(|u|_{\ult},|v|_{\ult})
\le |u|_{\ult}+|v|_{\ult}$.

\end{enumerate}
}
\vskip 0.5 cm

Here we will remark on this ultrametric as follows [M].

\vskip 0.5 cm

\begin{enumerate}

\item
If we define the distance $d(x,y)$ between points $x, y \in {\cal
A}_{[\beta]}\cup\{0\}$ by $d(x,y):= | \ |u-v|\ |_{\ult}$, the fifth
property in II${}_\beta$ satisfies (2), since the absolute value
$|u-v|$ belongs to ${\cal A}_{[\beta]}\cup\{0\}$. This metric induces
very a week topology.

\item
Since $x \in {\cal A}_{[\beta]}$ has
a finite value at $\beta\to\infty$, we have relation
\Be
\left.{|x|_{\ult}}\right|_{\bar \beta\sim \infty} \sim
\left.\exp( - \bar \beta (-(\log x)/\beta))\right|
_{\bar \beta \sim \beta \sim \infty}
=|x|^{  \bar \beta/\beta}|_{\bar \beta \sim \beta \sim \infty}.
\Ee
It may be regarded that $|x|_{\ult}
\sim |x|$, in heart, by synchronizing $\bar \beta$ and $\beta$.
$|x|_\ult$ is consist with  the
natural metric $|\cdot|$ in $\RR$.

\item
In this metric, we have the relation,
\Be
        | \sum_m x_m |_{\ult} = \ee^{-\bar \beta
        \min(\ordu(x_m)) }.
\Ee
This relation appears in the partition function at
$\bar \beta\sim\beta=1/T$, $T \to 0$ and
in the semi-classical path integral
$\bar \beta \sim
\beta=1/\hbar$, $\hbar\to 0$
[D, FH].
It means that {\it the classical regime appears as a
non-Archimedean valuation, which is an algebraic
manipulation}.
For example, a problem with
a minimal principle might be regarded
as a valuation of a quantum problem.

\end{enumerate}

These remarks shows that the ultrametric obtained from
the ultra-discrete is a very
natural object from physical and mathematical viewpoints.

\vskip 1.0 cm

{\centerline{\Large{\bf \S 3. Preliminary: $p$-adic Space}}}

\vskip 0.5 cm

In the ultrametric space theory, $p$-adic space is prototype. Hereafter
let us  consider the $p$-adic space. In this section, let us introduce
$p$-adic field $\QQ_p$ and $p$-adic integer $\ZZ_p$ for a prime number
$p$ [C, I, L, RTV, VVZ]. For a rational number $u \in \QQ$ which is
given by $u = \displaystyle{\frac{v}{w}} p^m$ ($v$ and $w$ are coprime
to the prime number $p$ and $m$ is an integer), we define a symbol
$[[u]]_p = p^m$. Here let us define  the $p$-adic valuation of $u$ as
a map from $\QQ$ to a set of integers $\ZZ$,
\Be
 \ordp(u):=  \log_p [[u]]_p, \ \mbox{ for } u \neq 0 ,
 \ \mbox{ and }
        \ \ordp(u):= \infty,\  \mbox{ for } u = 0. 
\Ee
This valuation has following properties (I${}_p$), which
is similar to I${}_\beta$ of $\ordu$,

\vskip 0.5 cm

{\large { \bf Proposition I${}_p$:}}
{\it
For $u,v \in \QQ$,
\begin{enumerate}

\item $ \ordp(u v ) = \ordp(u) + \ordp( v ) $.

\item $ \ordp(u + v ) \ge \min( \ordp(u) , \ordp(
v ))$.

 If $\ordp(u) \neq \ordp(v)$,
$ \ordp(u + v ) = \min( \ordp(u) , \ordp( v ))$.

\end{enumerate}
}

\vskip 0.5 cm

This property (I${}_p$-1) means that $\ordp$ is a homomorphism from the
multiplicative group $\QQ^\times$ of $\QQ$ to the additive group $\ZZ$.
The $p$-adic metric is given by $|v|_p = p^{-\ordp(v)}$, which has the
properties (II${}_p$);

\vskip 0.5 cm

{\large \bf Proposition II${}_p$:}{\it
For $u,v \in \QQ$,

\begin{enumerate}

\item if $ |v|_p =0 $, $v=0$.

\item $ |v|_p \ge0 $.

\item $ |v u |_p= |v|_p |u|_p $.

\item $|u + v|_p \le\max(|u|_p, | v|_p)
\le |u|_p+|v|_p$.

\end{enumerate}
}

\vskip 0.5 cm

The $p$-adic field $\QQ_p$ is given as a completion of $\QQ$ with
respect to this metric so that properties (I${}_p$) and (II${}_p$)
survive for $\QQ_p$.

It should be noted that the properties I${}_p$ and II${}_p$ are
essentially the same as those of I${}_\beta$ and II${}_\beta$. As a
property of $p$-adic metric, the convergence condition of series $
\sum_{m} x_m$ is identified with the vanishing condition of sequence
$|x_m|_p \to 0$ for $m \to \infty$ due to the relationship [C, L, VVZ],
\Be
      | \sum_{m} x_m |_{p} =
        \max |x_m|_p. 
\Ee
This property is related to (9).

Further we note  that the integer part of $\QQ_p$, $\ZZ_p :=\{ u \in
\QQ_p \ | \ \ordp(u)>0 \}$, is a {\it localized ring} and has only
prime ideals $\{ {0} \}$ and $p \ZZ_p$ [L]. $\ZZ_p$ can be also defined
by an inverse limit of the projective sequence [L],
\Be
        \ZZ/p\ZZ \gets \ZZ/p^2\ZZ \gets \ZZ/p^3 \ZZ \gets \cdots,
\Ee
or
\Be
        \ZZ_p := \lim_{\gets} \ZZ/ p^N \ZZ. 
\Ee
Thus there is a natural surjective ring homomorphism from $\ZZ_p$ to
$\ZZ/ p^N \ZZ$.

\vskip 1.0 cm

{\centerline{\Large{\bf \S 4. $p$-adic  Difference-Difference
Lotka-Volterra Equation}}}
 {\centerline{\Large{\bf and Its Applications }}}

\vskip 0.5 cm

Now let us define the $p$-adic difference-difference  Lotka-Volterra
equation for a $p$-adic series $\{ c_n^m\in \QQ_p \ |\ \ (n,m) \in
\Omega\times \ZZ\ \}$ ($p \neq 2$),
\Be
 \frac{c^{m+1}_n}{ c^m_n} = \frac{ 1 + \delta_p
 c^m_{n-1}} { 1 + \delta_p c^{m+1}_{n+1} },
\Ee
or
\Be
 {c^{m+1}_n}({ 1 + \delta_p c^{m+1}_{n+1} })
={ c^m_n}({ 1 + \delta_p
 c^m_{n-1}}) ,
\Ee
where $ \delta_p \in p\ZZ_p$ and $|\delta_p|_p<1$.
We proved that this equation has non-trivial solution
in [M].

In this article, we will give another proof. Due to (12) and (13),
there is a natural projection from $\ZZ_p$ to $\ZZ/p^N \ZZ$.
The equation (15) can be solved by module computations if
$c^0_n \in \ZZ_p$ for all $n \in \Omega$. Let us expand $c^m_n$
in $p$-adic space,
\Be
        c^m_n:={\alpha^m_n}^{(0)}
        +{\alpha^m_n}^{(1)}p+{\alpha^m_n}^{(2)}p^2+
                \cdots,
\quad
        {c^m_n}^{(N)}:=\sum_{i=0}^{N-1}{\alpha^m_n}^{(i)}p^i. 
\Ee
For simplicity,
we assume $\delta_p\equiv p$.
By comparing the coefficients of $p^{N}$ $(N=0,1,2,\cdots)$ in
the both sides in (15),
we can determine the time revolution iteratively:
\Bea
  {\alpha^{m+1}_n}^{(0)}&=&{\alpha^{m}_n}^{(0)}, \nonumber\\
 {\alpha^{m+1}_n}^{(1)}
      &=&\left({c^{m}_n}(1+p c^{m}_{n-1})-{c^{m+1}_n}^{(1)}\right)/p
      \quad    \mbox{module }p, \nonumber\\
     & &\cdots \nonumber\\
        {\alpha^{m+1}_n}^{(N)}
     &=&\left({c^{m}_n}(1+p c^{m}_{n-1})-
            {c^{m+1}_n}^{(N)}(1+p\ {c^{m+1}_{n+1}}^{(N)})\right)/p^N
            \quad  \mbox{module }p.
\Eea
This comparison means that we compute (15) in modulo $p^{N+1}$,
$(N=0,1,2,\cdots)$. If the initial state is given by $\{c_n^0 \equiv
{c_n^0}^{(N_0)}\}_{n \in \Omega}$ for a finite $N_0$, above
computations give all values of ${\alpha_n^m}^{(N)}$'s.
Then ${\alpha_n^m}^{(N)}$'s  vanish for ${n\in\Omega}$, $N>N_1$,
sufficient large $N_1$, and finite $m$.
It implies that (15) has non-trivial solutions in $p$-adic
space.

Using the natural projection from $\ZZ_p$ to $\ZZ/p^N \ZZ$, we have
solutions (17) of the equation (15) modulo $p^{N+1}$ when we fix $N$.
We give some examples in tables 1, 2, and 3 with periodic boundary
condition for $n$; $c^{m}_{n}\equiv c^{m}_{n+M}$ and $M=5$. These are
of $p=3$ and $p=5$ modulo $p^3$ $(N=2)$ cases and $p=7$ modulo $p^2$
$(N=1)$ case. Numerical computations show that they are also periodic
on time $m$ $c_n^m=c_n^{m+p^{N}}$. Since ${\alpha^{m}_n}^{(0)}$  is an
invariance, the excitations look localized.  We can find that there
appear several symmetries; in the tables $c^m_4$ oscillate with shorter
periods $p^{N-1}$, $c^m_n$ has point symmetry centerizing at $(n=1.5,
m=p^{N}/2)$ and so on. We regard that these are solutions of
Lotka-Volterra over rings $\ZZ/p^N \ZZ$. It is also noted that (15)
over the finite field ${\bf F}_p\equiv \ZZ/p\ZZ$ gives only trivial
solutions since  ${\alpha^{m+1}_n}^{(0)}$ is invariant. These facts
mean that {\it we can define soliton equation over rings beside finite
fields} [N, NM].

\vskip 1.0 cm

Here we will mention the relation of $p$-adic equation (15) to the
ultra-discrete system following [M]. As the $p$-adic
difference-difference Lotka-Volterra equation is well-defined, let us
consider the $p$-adic valuation of the equation (14) even though the
conserved quantities ${\alpha^{m+1}_n}^{(0)}$ might make the valuation
trivial. By letting $f^m_n := - \ordp( c^m_n )$ and $d_p:=-\ordp(
\delta_p)$, we have
\Be
f^{m+1}_n - f^m_n = \ordp( 1 + \delta_p c^m_{n-1})
- \ordp(1 + \delta_p c^m_{n-1}).
\Ee
For $f^{m}_{n}\neq -d_p$, (18) becomes
\Be
f^{m+1}_n - f^m_n = \max( 0 , f^m_{n-1} + d_p) -
\max( 0,  f^{m+1}_{n+1} + d_p).
\Ee
We emphasize that (19) has the same form as the ultra-discrete
difference-difference Lotka-Volterra equation (6) [M]. We should note
that $p$-adic valuation is a natural object in the $p$-adic number and
the family of rings $\{ \ZZ/p^N \ZZ\}$. This implies that the
ultra-discrete difference-difference system should be also studied from
the point of view of valuation theory [M].

As we finish this section, we will give a comment on a relation to
$q$-analysis. It is known that some of properties in the $q$ analysis
can be regarded as those in $p$-adic analysis by setting $q=1/p$ [VVZ].
We have correspondence among $p$, $q$ and $ \ee^{\beta}$ as [M],
\Be
\ee^{- \beta}  \Longleftrightarrow p
 \ (\beta \sim \infty),
\quad p \Longleftrightarrow 1/q, \quad
q \Longleftrightarrow  \ee^{\beta}\ (\beta \sim 0).
\Ee

\vskip 1.0 cm

{\centerline{\Large{\bf \S 5. Summary and Discussion}}}

\vskip 0.5 cm

We showed that the ultra-discrete limit should be regarded as a
non-Archimedean valuation following the previous report [M]. After we
constructed the ultrametric related to the ultra-discrete limit in (7),
we remarked its properties. Due to the remarks, it is interpreted that
the ultra-discrete limit is a very natural manipulation in ${\cal
A}_\beta \cup \{ {0} \}$.

Generally in the studies of the ultrametric space [LL], the $p$-adic
system is a prototype. Thus we have considered  $p$-adic soliton
equation following the previous report [M]. In fact the structure of
$p$-adic valuation of the $p$-adic difference-difference equation has
the same structure as the ultra-discrete difference-difference equation
for the case of Lotka-Volterra equation.

Further since $p$-adic field has a natural projection to a finite ring
$\ZZ/p^N \ZZ$, we have studied the Lotka-Volterra equation over the
finite ring. Due to the finiteness of system, we can give concrete
solutions of the equation. There remains a problem what is
integrability in the sense of $\ZZ/p^N \ZZ$ but the numerical
computations give regular results and beautiful symmetries of the
system. It is expected that there is a group governing this system.
This construction can be easily extended to a $p$-adic system related
to more general algebraic integer if the algebraic integer is a
principal domain. As the soliton theory in finite field is closely
related to the code theory [N, NM], the soliton over $\ZZ/p^N \ZZ$
might be also applied to the information theory [K]. Further as the
discrete heat equation can be described algebraically in the
introduction, we wish, in future, to express the equation over $\ZZ/p^N
\ZZ$ more algebraically.

Finally we comment upon an open problem. The non-Archimedean valuation
theory is associated with the measure theory or non-standard statistics
[LL] and renormalization theory [RTV]. On the other hand, soliton
theory is connected with statistical system  and statistical mechanics
[So]. Thus we have a question whether both non-standard statistics and
soliton theory have more directly relation.

\vskip 1.0 cm

{\centerline{\Large{\bf  Acknowledgment}}}

\vskip 0.5 cm

I would like to thank Prof. K. Sogo for giving me
chances to use his personal computer.


\vskip 1.0 cm
\centerline{\Large{\bf Reference}}

\begin{enumerate}

\item[[C]]  J.W.S. Cassels, {\it Lectures on
Elliptic Curves }, (1991, Cambridge Univ. Press,
 Cambridge).

\item[ [ D]] P.~A.~M.~Dirac,
{\it The Principles of Quantum Mechanics, forth edition},
(1958, Oxford, Oxford).

\item[ [ FH]]  R.~P.~Feynman and A.~R.~Hibbs,
{\it Quantum Mechanics and Path Integral},
(1965, McGraw-Hill, Auckland).

\item[ [HT] ]    R. Hirota and S. Tsujimoto,
 {\it  Conserved Quantities of a Class
of Nonliear Difference-Difference Equations },
J. Phys. Soc. Jpn.{\bf  64 } (1995)
3125-3127.

\item[ [Ic1] ] T. Ichikawa,
 {\it The universal periods of curves and the Schottky problem },
Comp. Math. {\bf  85}  (1993)
1-8.

\item[ [Ic2] ]   ----,
 {\it $p$-adic theta functions and solutions of the KP
hierarchy },
Comm. Math. Phys. {\bf 176} (1996)
383-399.

\item[ [Ic3] ]  ----,
 {\it  Schottky uniformarization theory on Riemann surfaces
and Mumford curves of infinite genus }
J. reine angew. Math. {\bf 486} (1997)
  45-68.

 \item[ [I] ]     K. Iwasawa, {\it  Algebraic
Function Theory}, ( 1952, Iwanami, Tokyo) {\it japanese}.

\item[ [K] ] N. Koblitz,
{\it Algebraic Aspects of Cryptography}
 (1998, Springer, Berlin).

\item[ [L] ] K. Lamotke (ed.), {\it Zahlen}
 (1983, Springer, Berlin).

 \item[ [LL] ]   A.~J.~Lemin and V.~A.~Lemin,
 {\it  On a universal ultrametric space },
Topology and its Applications {\bf  103 }  (2000)
339-345.

\item[ [M] ]  S. Matsutani,
 {\it   $p$-adic Difference-Difference
 Lotka-Volterra Equation and Ultra-Discrete Limit},
Int. J. Math. and Math. Sci.
(2001) to appear, (solv-int/9906011).

 \item[ [N] ]  Y.~Nakamura,
{\it  The BCH-Goppa decoding
as a moment problem and a tau function over
finite fields },
Phys. Lett. A {\bf  223 } (1996)
75-81.

 \item[ [NM] ]   Y.~Nakamura and A. Mukaihira,
{\it
Dynamics of the finite Toda molecule over
finite fields and a decoding algorithm
},
Phys. Lett. A {\bf  249 } (1998)
295-302.

 \item[ [RTV] ]     R. Rammal, G. Toulouse and M. A. Virasoro,
{\it  Ultrametricity for physicists },
 Rev. Mod. Phys.{\bf  58}  (1986)
765-788.

 \item[ [PTVF] ]   W.~H.~Press, S.~A.~Teukolsky,
W.~T.~Vetterling and B.~P.~Flannery
{\it  Numerical Recipes in Fortran 77: The Art of
Scientific Computing (Vol.~1 of Fortran Numerical Recipes), 2nd ed.},
(1996, Cambridge Univ. Press, Cambridge).

\item[ [S] ]  M. Sato,
 {\it Soliton Equation and Universal Grassmannian Manifold },
 Lecture note, noted by M.~Noumi, ({\it japanese})
(1984, Sophia Univ, Tokyo).

 \item[ [So] ]    K. Sogo,
 {\it
Time-Dependent Orthogonal Polynomials and Theory
of Soliton - Applications to Matrix Model,
Vertex Model and Level Statistics },
 J. Phys. Soc. Jpn {\bf  62 } (1993)
1887-1894.

 \item[ [T] ]  D. Takahashi,
{\it  Ultra-discrete Toda Lattice Equation -A Grandchild of
 Toda- },
International Symposium, Advances in soliton theory
and its applications: The 30th anniversary of
the Toda lattice (1996)
36-37.

 \item[ [TS] ]     D. Takahashi and J. Satsuma,
 {\it  A Soliton Cellular Automaton },
J. Phys. Soc. Jpn.{\bf  59} (1990)
3514-3519

 \item[ [TTMS] ]    T. Tokihiro, D. Takahashi, J.
Matsukidaira and J. Satsuma,
{\it  From
Soliton Equations to Integrable Cellular Automata
through a Limiting Procedure },
  Phys.Rev.Lett.
{\bf  76} (1996)  3427-3250.

 \item[ [VVZ] ]    V.S.Vladimirov, I.V.Volvich and
E.I.Zelenov, {\it  $P$-adic Analysis and Mathematical
Physics}  (1994, World Scientific, Singapore).
\end{enumerate}

\newpage

$$
\centerline{
\vbox{
        \baselineskip =10pt
        \tabskip = 1em
        \halign{&\hfil#\hfil \cr
    \multispan7 \hfil Table 1: modulo $3^3$, $\delta_p=3$ \hfil \cr
        \noalign{\smallskip}
        \noalign{\hrule height0.8pt}
        \noalign{\smallskip}
& $m\ \backslash  \ n$ &\strut\vrule& 0  & 1   & 2 & 3 & 4 & 5  \cr
\noalign{\smallskip}
\noalign{\hrule height0.3pt}
\noalign{\smallskip}
& $0$ & \strut\vrule&  1 & 2 & 2 & 1 & 1 & 1\cr
& $1$ & \strut\vrule&  7 & 23 & 26 & 22 & 10 & 7\cr
& $2$ & \strut\vrule&  13 & 8 & 14 & 16 & 10 & 13\cr
& $3$ & \strut\vrule&  19 & 11 & 20 & 10 & 1 & 19\cr
& $4$ & \strut\vrule&  25 & 5 & 17 & 4 & 10 & 25\cr
& $5$ & \strut\vrule&  4 & 17 & 5 & 25 & 10 & 4\cr
& $6$ & \strut\vrule&  10 & 20 & 11 & 19 & 1 & 10\cr
& $7$ & \strut\vrule&  16 & 14 & 8 & 13 & 10 & 16\cr
& $8$ & \strut\vrule&  22 & 26 & 23 & 7 & 10 & 22\cr
& $9$ & \strut\vrule&  1 & 2 & 2 & 1 & 1 & 1\cr
\noalign{\smallskip}
        \noalign{\hrule height0.8pt}
}
}
}
$$

\newpage
$$
\centerline{
\vbox{
        \baselineskip =10pt
        \tabskip = 1em
        \halign{&\hfil#\hfil \cr
     \multispan7 \hfil Table 2: modulo $5^3$, $\delta_p=5$ \hfil \cr
        \noalign{\smallskip}
        \noalign{\hrule height0.8pt}
        \noalign{\smallskip}
& $m\ \backslash \ n$ &\strut\vrule& 0  & 1   & 2 & 3 & 4 & 5  \cr
\noalign{\smallskip}
\noalign{\hrule height0.3pt}
\noalign{\smallskip}
& $0$ & \strut\vrule&   1 & 2 & 2 & 1 & 1 & 1\cr
& $1$ & \strut\vrule&    96 & 117 & 37 & 106 & 26 & 96\cr
& $2$ & \strut\vrule&    16 & 7 & 97 & 36 & 101 & 16\cr
& $3$ & \strut\vrule&    11 & 47 & 57 & 41 & 101 & 11\cr
& $4$ & \strut\vrule&    81 & 112 & 42 & 121 & 26 & 81\cr
& $5$ & \strut\vrule&    101 & 77 & 52 & 26 & 1 & 101\cr
& $6$ & \strut\vrule&    71 & 67 & 87 & 6 & 26 & 71\cr
& $7$ & \strut\vrule&    116 & 82 & 22 & 61 & 101 & 116\cr
& $8$ & \strut\vrule&    111 & 122 & 107 & 66 & 101 & 111\cr
& $9$ & \strut\vrule&    56 & 62 & 92 & 21 & 26 & 56\cr
& $10$ & \strut\vrule&    76 & 27 & 102 & 51 & 1 & 76\cr
& $11$ & \strut\vrule&    46 & 17 & 12 & 31 & 26 & 46\cr
& $12$ & \strut\vrule&    91 & 32 & 72 & 86 & 101 & 91\cr
& $13$ & \strut\vrule&    86 & 72 & 32 & 91 & 101 & 86\cr
& $14$ & \strut\vrule&    31 & 12 & 17 & 46 & 26 & 31\cr
& $15$ & \strut\vrule&    51 & 102 & 27 & 76 & 1 & 51\cr
& $16$ & \strut\vrule&    21 & 92 & 62 & 56 & 26 & 21\cr
& $17$ & \strut\vrule&    66 & 107 & 122 & 111 & 101 & 66\cr
& $18$ & \strut\vrule&    61 & 22 & 82 & 116 & 101 & 61\cr
& $19$ & \strut\vrule&    6 & 87 & 67 & 71 & 26 & 6\cr
& $20$ & \strut\vrule&    26 & 52 & 77 & 101 & 1 & 26\cr
& $21$ & \strut\vrule&    121 & 42 & 112 & 81 & 26 & 121\cr
& $22$ & \strut\vrule&    41 & 57 & 47 & 11 & 101 & 41\cr
& $23$ & \strut\vrule&    36 & 97 & 7 & 16 & 101 & 36\cr
& $24$ & \strut\vrule&    106 & 37 & 117 & 96 & 26 & 106\cr
& $25$ & \strut\vrule&    1 & 2 & 2 & 1 & 1 & 1\cr
\noalign{\smallskip}
        \noalign{\hrule height0.8pt}
}}
}
$$

\newpage
$$
\centerline{
\vbox{
        \baselineskip =10pt
        \tabskip = 1em
        \halign{&\hfil#\hfil \cr
    \multispan7 \hfil Table 3: modulo $7^2$, $\delta_p=7$ \hfil \cr
        \noalign{\smallskip}
        \noalign{\hrule height0.8pt}
        \noalign{\smallskip}
& $m\ \backslash  \ n$ &\strut\vrule& 0  & 1   & 2 & 3 & 4 & 5  \cr
\noalign{\smallskip}
\noalign{\hrule height0.3pt}
\noalign{\smallskip}
& $0$ & \strut\vrule&     1 & 2 & 2 & 1 & 1 & 1\cr
& $1$ & \strut\vrule&     43 & 37 & 16 & 8 & 1 & 43\cr
& $2$ & \strut\vrule&     36 & 23 & 30 & 15 & 1 & 36\cr
& $3$ & \strut\vrule&     29 & 9 & 44 & 22 & 1 & 29\cr
& $4$ & \strut\vrule&     22 & 44 & 9 & 29 & 1 & 22\cr
& $5$ & \strut\vrule&     15 & 30 & 23 & 36 & 1 & 15\cr
& $6$ & \strut\vrule&     8 & 16 & 37 & 43 & 1 & 8\cr
& $7$ & \strut\vrule&     1 & 2 & 2 & 1 & 1 & 1\cr
\noalign{\smallskip}
        \noalign{\hrule height0.8pt}
\noalign{\smallskip}
}}
}
$$
\enddocument